\documentclass[12pt,english]{article}
\usepackage[T1]{fontenc}
\usepackage[a4paper]{geometry}
\usepackage{babel}
\usepackage{textcomp}
\usepackage[unicode=true]
 {hyperref}
 \usepackage{amsthm}
 \usepackage{amssymb}
 \usepackage{pgfplots}
   	\usepgfplotslibrary{external}
  \pgfplotsset{compat=newest}
  \usetikzlibrary{plotmarks}
  \usetikzlibrary{arrows.meta}  
  \usetikzlibrary{patterns}
  \usepgfplotslibrary{patchplots}
  \usepackage{grffile}
  \usepackage{amsmath}
  \usepackage{tikz}
\usepackage{graphicx}
\usepackage{subfig}
\usepackage{stmaryrd}
\DeclareMathOperator{\e}{e}
\DeclareMathOperator{\tr}{tr}

\usepackage[numbers]{natbib}

\makeatletter
\newcommand{\lyxaddress}[1]{
	\par {\raggedright #1
	\vspace{1.4em}
	\noindent\par}
}

\@ifundefined{date}{}{\date{}}
\makeatother
\usepackage{times}
\begin{document}
\title{Analytical and experimental results: creation of a duffing acoustic nonlinear oscillator at low amplitudes}
\author{M. Morell$^{1*}$, M. Collet$^{2}$, E. Gourdon$^{1}$, A. Ture Savadkoohi$^{1}$, E. De Bono$^{3}$}
\maketitle

\lyxaddress{$^{1}$ ENTPE, Ecole Centrale de Lyon, CNRS, LTDS, UMR5513, 69518 Vaulx-en-Velin, France}

\lyxaddress{$^{2}$ CNRS, Ecole Centrale de Lyon, ENTPE, LTDS, UMR5513, 69130 Ecully, France}

\lyxaddress{$^{3}$ Ecole Centrale de Lyon, CNRS, ENTPE, LTDS, UMR5513, 69130 Ecully, France}

\lyxaddress{$^{*}$ Corresponding author, maxime.morell@entpe.fr}

\begin{abstract}
Nonlinear dynamics have long been exploited in order to damp vibrations in solid mechanics. The phenomenon of irreversible energy transfer from a linear primary system to a nonlinear absorber has driven great attention to the optimal design of vibration absorbers both for stationary and transient regimes. Recently, the same principle has also been targeted in acoustics for the absorption of sound waves at high excitation amplitudes. Meanwhile, acoustic absorption by electro-active means has found great success for noise reduction in the linear regime. This study uses a method allowing the design of nonlinear resonators at amplitudes that typically induce linear behaviors. This research proposes an analytical study of the implementation of the duffing equation as a nonlinear electroacoustic resonator coupled to an acoustic mode of a tube. Experiments are carried out and compared to the analytical results. The experimental implementation is done using a real-time-based algorithm retrieving the measured pressure from a microphone and giving the electrical current to send to a loudspeaker as an output thanks to a Runge-Kutta-like algorithm.
\end{abstract}

\section{Introduction}
The control of sound and vibrations has been studied for years in various domains, such as aeronautics \cite{Morse1939}. In the field of acoustics, multiple kinds of devices have been considered, from acoustic foams to active noise-cancellation devices. Passive absorbing materials are the most commonly used way to reduce sound levels as these materials are very efficient for frequencies greater than 1000 Hz \cite{Delany1970}, while they struggle at lower frequencies \cite{Champoux1991}. The use of active devices consists of creating a wave in phase opposition to the incident wave, resulting in the cancellation of wave energy \cite{Lueg1934,Miljkovic2016}. Nevertheless, this concept requires large amounts of energy, as the created wave amplitude should be equal to the incident wave one \cite{Guicking2009,Gardonio2004,Gardonio2004a,Bianchi2004}. Moreover, in more than 1 dimension, the energy canceling due to the wave superposition is efficient only in local spatial zones. Passive resonators do not use additional energy to be efficient but are specifically designed to reduce sound pressure at their resonance frequency. For low frequencies, their design can be massive, and they can not be adjusted to the frequency of incident waves such as the Helmholtz resonator. To create resonators that can be tuned, the use of loudspeakers controlled by electrical current has been found as being a solution by Olson and May \cite{Olson1953}. This method named Impedance Control (IC) consists of adapting the apparent acoustic impedance of the loudspeaker thanks to microphones \cite{Guicking1984a, Furstoss1997, Thenail1997}. Indeed, an electrical current is sent to the loudspeaker coil based on the measured pressure and on calculations done with a processor to change the loudspeaker parameters \cite{Karkar2019,DeBono2022}. In the acoustic field, these resonators remained in their linear domain, and the advantages of nonlinearities \cite{Gatti2019} have not been fully employed yet due to the high activation threshold of the nonlinearity, unlike the field of mechanics \cite{Vakakis2009}. Nevertheless, Bellet \cite{Bellet2010} excited a visco-elastic membrane at very high amplitudes (above 148 dB Sound Pressure Level (SPL)) to control a tube acoustic mode. Additionally, Gourdon \cite{Gourdon2018,Vargas2017} activated the nonlinear regimes of a Helmholtz resonator above 138 dB SPL. Guo \cite{Guo2020} carried out the creation of a polynomial nonlinear electro-acoustic resonator at low excitation amplitudes (close to 95 dB) using an additional microphone placed in the back cavity of the loudspeaker. The additional microphone permits to add a polynomial nonlinear current to a frequency-based linear current. In this study, a nonlinear electroacoustic resonator is created at low excitation amplitudes using the real-time-based method presented and validated in \cite{DeBono2023}, which does not need additional microphones. This method can create polynomial and non-polynomial behaviors. This research features the programming of a cubic restoring force function to create a duffing resonator. The nonlinear electroacoustic resonator is coupled to an acoustic mode of a tube. In section \ref{Sec1}, the system under consideration is presented and the programming strategy is briefly introduced. The analytical model presented in section \ref{Sec1} is studied thanks to analytical methods in section \ref{Sec2}. The limits of the hypothesis of such methods are highlighted. Experimental results are presented in Section \ref{Sec3}, with comparisons to the analytical results.
\section{System under consideration}
\label{Sec1}
\subsection{The nonlinear electroacoustic resonator}
An electroacoustic resonator is composed of an actuator (a loudspeaker) collocated to sensors (one or many microphones) and equipped with a processor running an algorithm that takes the measured pressure as an input and gives the electrical current to send into the loudspeaker coil as an output.
Nonlinear responses in resonators are commonly initiated at high amplitudes of motion. This research suggests employing the method presented and detailed by De Bono et al. \cite{DeBono2023} to create a nonlinear electroacoustic resonator (ER) behaving as a duffing oscillator within the displacement and pressure amplitude ranges corresponding to the linear regime of the loudspeaker. It consists of adding a programmed additional force to the passive resonator in order to drive its behavior from its inherent behavior to a desired behavior. This means that the resonator can be digitally programmed.\\
The experimental application of the algorithm assumes that the operational frequencies are close to the frequencies of the loudspeaker's first mode. Additionally, it presupposes that the operational amplitudes remain within the linear regime range of the loudspeaker. Given these assumptions, along with the consideration that the frequencies and amplitudes are sufficiently low to treat the membrane as rigid, the loudspeaker can be approximated using the classical mass-spring-damper model described by the following equation (\ref{DynER}):
\begin{equation}\label{DynER}
    M_{0} \ddot u(t) + R_{0} \dot u(t) + K_{0} u(t) = p(t)S_d -Bli(t)
\end{equation}
where $M_0$, $R_0$, and $K_0$ represent the eigen parameters of the loudspeaker first mode when the electrical current $i$ is set to zero. $Bl$ is the force factor of the solenoid of the loudspeaker, with $B$ the magnetic field produced by the permanent magnet, and $l$ the length of the solenoid. As a result, $Bli$ describes the Laplace force through which the control is applied. $u$ stands for the relative displacement of the membrane regarding its resting position, and $\dot \bullet$ indicates the time derivative of the considered variable. $p$ denotes the pressure of the incident wave applied on the membrane of the loudspeaker, and $S_d$ the effective area of the membrane. Equation (\ref{DynER}) describes the inherent behavior of the loudspeaker.\\

The goal of the electrical current is to introduce the Laplace force in order to modify the intrinsic response of the loudspeaker into a desired response specified in equation (\ref{DesER}):
\begin{equation}\label{DesER}
   M_{t}\ddot u_t(t)+R_{t} \dot u_t(t) + K_{t} u_t(t) + F(t,u_t(t),\dot u_t(t), \ddot u_t(t)) = p(t)S_d
\end{equation}
where $M_t$, $R_t$, and $K_t$ represent the targeted eigen parameters of the loudspeaker first mode. $F$ is a nonlinear function, such as a mass, damper, or stiffness force. These parameters can be chosen in the programming. The variable $u_t$ stands for the targeted displacement and is induced by the measured pressure $p$ and by the choice of these parameters through the numerical resolution of equation (\ref{DesER}) at each time step.\\
Once the numerical resolution of equation (\ref{DesER}) yields the targeted displacement $u_t$, velocity $\dot{u}_t$, and acceleration $\ddot{u}_t$, the driving parameter of the Laplace force $i$ can be calculated utilizing equation (\ref{Id_elec}), assuming $u = u_t$:
\begin{equation}\label{Id_elec}
    i(t)=\frac{S_d}{Bl}\Bigg(p(t) - \Big(\frac{M_0}{S_d} \ddot u_t(t) + \frac{R_0}{S_d} \dot u_t(t) + \frac{K_0}{S_d} u_t(t)\Big)\Bigg)
\end{equation}
The controlling variable of the programming is the electrical current $i$ and is calculated at each time step. Upon closer examination of this variable and substituting $i$ from equation (\ref{Id_elec}) into equation (\ref{DynER}), it yields:
\begin{equation}\label{Expl}
    M_0(\ddot u - \ddot u_t) + R_0(\dot u - \dot u_t) + K_0(u - u_t) = 0
\end{equation}
Equation (\ref{Expl}) highlights that the force induced with $i$ is applied to enforce the behavior carried out in the calculations of $u_t$, $\dot u_t$ and $\ddot u_t$. The objective of the algorithm is to drive the left-hand side of Equation (\ref{Expl}) to zero. When this occurs, the targeted behavior is attained. It can be seen that the Laplace force introduces an additional force applying on the loudspeaker membrane to modify its displacement to become the targeted displacement, which is a duffing behavior displacement when $F_{NL}(u_t)=\pm K_t\beta_{NL}u_t^3(t)$.\\
The internal parameters of the loudspeaker are measured and presented in Table \ref{table_ER_0}:
\begin{table}[ht!] 
	\centering
	\begin{tabular}{c|c}
		$M_0$     &$3.89\times10^{-4}\ \text{kg}$\\
		$R_0$   &$2.63\times10^{-1}\ \text{kg}.\text{s}^{-1}$\\
		$K_0$     &$4.34\times10^{3}\ \text{kg}.\text{s}^{-2}$\\
		$Bl/S_d$  &$136.7\ \text{kg}.\text{m}^{-1}.\text{A}^{-1}.\text{s}^{-2}$
	\end{tabular}
	\caption{Values of the modal parameters of the first mode of the loudspeaker when the current is set to zero.}
	\label{table_ER_0}
\end{table}
\subsection{The experiment}

\begin{figure}[htp] 
	\centering
	\includegraphics{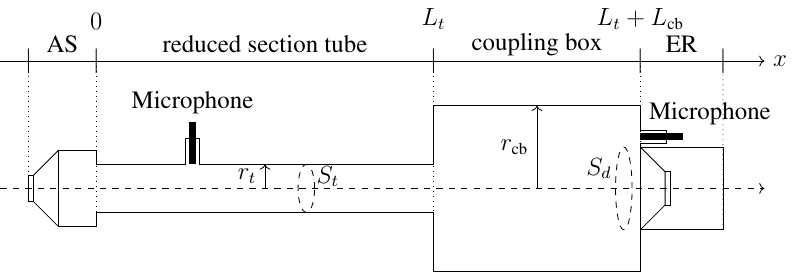}
	\caption{Scheme of the experimental set-up}
	\label{Schema_exp}
\end{figure}

Based on the previously outlined programming, the objective is to control an acoustic mode of a tube with a nonlinear resonator, ensuring that the amplitudes are sufficiently low for the air propagation to be linear and the loudspeaker to operate within its linear regime.\\
The presented experiment is built upon the framework introduced by Bellet \cite{Bellet2010} and Gourdon \cite{Gourdon2018}. The model for this experiment has been thoroughly explored in earlier works \cite{Bellet2010, Gourdon2018, Cochelin2006}, and its detailed development is not done in this study.\\
The experiment is composed of a circular tube of length $L_t$ and radius $r_t$. At one of its ends, an external loudspeaker is placed as an acoustic source (AS) for excitation of the system, modeled by an incident pressure $p_{ls}$. To be able to model the coupling between the ER and the acoustic mode of the tube, a circular coupling box of dimensions specified by length $L_{cb}$ and radius $r_{cb}$ is positioned at the other end of the tube. A coupling box consists of connecting a wider tube to the considered tube, permitting the establishment of a relation between the pressure and the volume variations caused by the displacement of the air particles of the reduced section tube and the displacement of the membrane of the ER. Let us set the axis passing through the center of the cross sections of the tube associated with the variable $x\in[0,L_t]$. The dimensions of the experiment are presented in Table \ref{Table_Value}. A schematic representation of the experiment is depicted in Figure \ref{Schema_exp}.\\
\begin{table}[ht!]
    \centering
    \begin{tabular}{c|c}
    $r_t$     &$1.45\times10^{-2}\ \text{m}$\\
    $S_t$   &$6.61\times10^{-4}\ \text{m}^2$\\
    $L_t$     &$0.204\ \text{m}$\\
    $r_{\text{cb}}$  &$5.0\times10^{-2}\ \text{m}$\\
    $L_{\text{cb}}$    &$0.125\ \text{m}$\\
    $S_d$   &$1.3\times10^{-3}\ \text{m}^2$
    \end{tabular}
    \caption{Values of the dimension of the experiment}
    \label{Table_Value}
\end{table}
The experiment has been designed in order to be modeled using a two-degree-of-freedom system. Considering that the variables of time $t$ and space $x$ of the solution can be decoupled, the governing equations of the system are:
\begin{equation}\label{EqSystem}
\left\{
    \begin{aligned}
        &m_a\ddot u_a(t) +c_a \dot u_a(t) + k_a u_a(t)+\frac{\gamma}{\alpha}\big(u_a(t) - \alpha u_m(t)\big)=-S_t p_{\text{ls}}(t)\\
        &M_t \ddot u_m(t)+ R_t \dot u_m(t) + K_t u_m(t) + F_{\text{NL}}\big(u_m(t)\big)+\gamma \big(\alpha u_m(t) - u_a(t)\big)=0 
    \end{aligned}
\right.
\end{equation}
where $\alpha$ is defined as the ratio of the sections of the tube and of the coupling box $\alpha=S_d/S_t$. The coefficients $m_a$, $c_a$ and $k_a$ stand for the modal parameters of the first acoustic mode of the tube, linked to $u_a$ the modal coordinates of the first acoustic mode of the tube. These coefficients are obtained from the variational formulation of the one-dimensional acoustic propagation equation, and from employing a Rayleigh-Ritz method:
\begin{equation}
    u_{air}(x,t)=u_a(t)\Phi_1(x)=u_a(t)\Big(-\cos{\frac{\pi x}{L_t}}\Big)
\end{equation}
where $\Phi_1$ represents the shape function (which is also the shape mode in this particular case). The parameter $\gamma$ stands for the stiffness of the spring coupling the acoustic mode of the tube and the ER:
\begin{equation}\label{CouplageFaible}
    \gamma=k_bS_tS_d
\end{equation}
with $k_b=\frac{\rho_0 c_0^2}{V_{\text{cb},0}}$. The parameters $\rho_0$ and $c_0$ denote the density of air and sound speed in air under standard conditions, while $V_{\text{cb},0}$ stands for the volume of the coupling box cavity at its resting state.
One can notice that the coupling between the two equations of system (\ref{EqSystem}) representing the motion of the acoustic first mode of the tube and the motion of the ER is non-reciprocal. This non-reciprocity is due to the distinct areas of the tube and the ER.
The parameters in the system of equations (\ref{EqSystem}) are either directly measured or estimated using the model. The estimated parameters are those that do not require experimental measurements, they are calculated based on the dimensions of the experiment. These parameters are presented in Table (\ref{Param_calculated}).
\begin{table}[ht!]
    \centering
    \begin{tabular}{c|c}
        $m_a$ & $8.22\times 10^{-5}\ \text{kg}$\\
        $k_b$ & $1.46\times 10^{8}\ \text{kg}.\text{m}^{-4}.\text{s}^{-2}$\\
        $\gamma$ & $125.54\ \text{kg}.\text{s}^{-2}$\\
    \end{tabular}
    \caption{Parameters obtained by calculations}
    \label{Param_calculated}
\end{table}
The measured parameters are directly estimated using experimental measurements. These parameters can rely on calculated parameters as the modal mass $m_a$ and are presented in Table \ref{Param_measured}.\\
\begin{table}[ht!]
    \centering
    \begin{tabular}{c|c}
        $f_0$ & $569.10 \pm 0.26\ \text{Hz}$\\
        $c_a$ & $ 3.65\times10^{-2} \pm 2.21\times10^{-5}\ \text{kg}.\text{s}^{-1}$ \\
        $k_a$ & $ 1014.40 \pm 0.95\ \text{kg}.\text{s}^{-2}$ 
    \end{tabular}
    \caption{Parameters obtained by measurements}
    \label{Param_measured}
\end{table}
The coupled equations (\ref{EqSystem}) allow an analytical study to be led using a perturbation approach.\\
\section{Method and Analytical study}
\label{Sec2}
The analytical method chosen to study such a system is the Multiple Scale Method (MSM), coupled to the complex variables of Manevitch \cite{Manevitch2001}. This method has been extensively employed for the study of nonlinear dynamical systems \cite{Lamarque2011,Hurel2019,TureSavadkoohi2016a,Labetoulle2022,DaSilveiraZanin2022, Bitar2020}.
\subsection{The standard form system}
The system under consideration is governed by system of equations (\ref{EqSystem}). This is a two-degree of freedom system. Let us consider the system rearranged as in \cite{TureSavadkoohi2016a}:
\begin{equation}\label{Eq_stand}
\left\{
    \begin{aligned}
        &u_a'' + \varepsilon \xi_a u_a' + u_a + \varepsilon \gamma_0 (u_a - \alpha u_m)= -\varepsilon f\sin{(\nu \tau)}\\
        &\varepsilon \Big(u_m'' +  \xi_m u_m' + K_Lu_m + K_{NL}u_m^3 + \gamma_0 \mu (\alpha u_m-u_a)\Big)=0
    \end{aligned}
\right.
\end{equation}
where $\varepsilon<<1$ and it has been set that $\bullet'=\frac{\partial}{\partial \tau}$ is a time differential operator regarding to the time $\tau$ defined as:
\begin{equation}
    \tau=\omega_0t=\sqrt{\frac{k_a}{m_a}}t
\end{equation}
and with $\nu$ the normalized pulsation of the excitation pulsation $\Omega$ defined as depending on $\sigma$ the detuning parameter, meaning that the solution is studied around the resonance of the system:
\begin{equation}
    \nu=\frac{\Omega}{\omega_0}=1+\sigma\varepsilon
\end{equation}
The MSM approach is introduced here by the decomposition of the time $\tau$ in new time scales as:
\begin{equation}\label{temps}
    T_k=\varepsilon^k\tau\ ; \     k\in\llbracket 0~;~ N \rrbracket.   
\end{equation}
Let us introduce the complex variables of Manevitch $(\varphi_a,\varphi_m)\in \mathbb{C}^2$ and are functions of new time scales $T_k\in\mathbb{R}$ \cite{Manevitch2001}, with $i^2=-1$:
\begin{equation}
    \left\{
    \begin{aligned}
        &u_a'+i \nu u_a=\varphi_a(T_0,T_1,T_2,...) \e ^{i\nu\tau}\\
        &u_m'+i \nu u_m=\varphi_m(T_0,T_1,T_2,...) \e ^{i\nu\tau}\\
    \end{aligned}
\right.
\end{equation}

Additionally, a Galerkin method is used to reduce the order of the problem to the first harmonic of the solution, assuming that the solution may be decomposed as a Fourier series, \textit{i.e.} a sum of mono-harmonic waves. Let us consider a general function $z: \mathbb{C}^n\ \longrightarrow \mathbb{C}^n$, with $n$ a natural finite number, and depending of a vector $\mathbf{w}$:
\begin{equation}
    z(\mathbf{w})=\sum_k Z_k(\mathbf{w})\e^{ik\nu t}
\end{equation}
where:
\begin{equation}
    Z_k(\mathbf{w})=\frac{\nu}{2\pi}\int_0^{\frac{\nu}{2\pi}}z(\mathbf{w})\e^{-ik\nu t} dt
\end{equation}
Naturally, the Galerkin procedure to truncate the function $z$ to its first harmonic consists of applying the scalar product of $\mathcal{L}^2(\mathbb{C})$, which is the following function $G:\mathbb{C}^n \longrightarrow \mathbb{C}^n$ to project onto the first harmonic:  
\begin{equation}
    G(z(\mathbf{w}))=\frac{\nu}{2\pi}\int_0^\frac{2\pi}{\nu} z(\mathbf{w})\e^{-i\nu t} dt
\end{equation}
By introducing the complex variables of Manevitch, and applying the Galerkin procedure supposing that $\varphi_a$, $\varphi_m$ and their complex conjugate (denoted by $\bullet^*$) $\varphi_a^*$, $\varphi_m^*$ are independent of $T_0=\tau$ (this hypothesis is verified after the fact), the system of equations (\ref{Eq_stand}) reads:
\begin{equation}\label{Eq_Galerkin}
\left\{
    \begin{aligned}
        &\frac{\text{d}\varphi_a}{\text{d}\tau}+i\nu\frac{\varphi_a}{2}+\varepsilon\xi_a\frac{\varphi_a}{2} + \frac{\varphi_a}{2i\nu}+\varepsilon \frac{\gamma_0}{2i\nu}(\varphi_a-\alpha\varphi_m)=\varepsilon\frac{-f}{2i}\\
        &\varepsilon\Big(\frac{\text{d} \varphi_m}{\text{d} \tau}+i\nu\frac{\varphi_m}{2}+\xi_m\frac{\varphi_m}{2} +K_L\frac{\varphi_m}{2i\nu} - K_{NL}\big(\frac{3i}{8\nu^3}\varphi_m^2\varphi_m^*\big)+\frac{\gamma_0\mu}{2i\nu}(\alpha\varphi_m-\varphi_a)\Big)=0 
    \end{aligned}
\right.
\end{equation}

Equation (\ref{temps}) yields the definition of the time derivative regarding $\tau$:
\begin{equation}
    \frac{\text{d}}{\text{d} \tau}=\sum_{k=0}^{N}\varepsilon^k \frac{\partial}{\partial T_k}
\end{equation}
System of equations (\ref{Eq_Galerkin}) yields equations at each order of $\varepsilon$. In our study, the equations corresponding to the different orders $O(\varepsilon^k)$ with $k\in\llbracket 0~;~ N \rrbracket$ are considered. The orders correspond to the time scale considered and as a result to the fast and slow system dynamics. 

\subsection{Fast system dynamics}
The system at $O(\varepsilon^0)$ is written with the consideration that $\nu=1+\varepsilon\sigma$:
\begin{equation}\label{eps0}
O(\varepsilon^0):
    \left\{
    \begin{aligned}
        &\frac{\partial \varphi_a}{\partial T_0}+i\frac{\varphi_a}{2}+ \frac{\varphi_a}{2i}=0\\
        &\frac{\partial \varphi_m}{\partial T_0}+i\frac{\varphi_m}{2}+\xi_m\frac{\varphi_m}{2} +K_L\frac{\varphi_m}{2i} - K_{NL}\big(\frac{3i}{8}\varphi_m^2\varphi_m^*\big)+\frac{\gamma_0\mu}{2i}(\alpha\varphi_m-\varphi_a)=0 
    \end{aligned}
\right.
\end{equation}
The first equation of the system yields that:
\begin{equation}
    \frac{\partial \varphi_a}{\partial T_0}=0
\end{equation}
which confirms the previously assumed hypothesis "$\varphi_a$ independent of $T_0$". The second equation of the system can be studied in the case of an asymptotic state, when $T_0 \longrightarrow \infty$. It gives that:
\begin{equation}\label{Hyp_asympt}
    \frac{\partial \varphi_m}{\partial T_0}\longrightarrow 0
\end{equation}
As a result, the system of equations reads:
\begin{equation}\label{SIM}
    i\frac{\varphi_m}{2}\Big(1-K_L-\frac{3}{4}K_{NL}|\varphi_m|^2-\alpha \gamma_0 \mu -i\xi_m\Big)+i\gamma_0 \mu \frac{\varphi_a}{2}=0
\end{equation}
We define the map $\mathcal{H}$ as it follows:
\begin{equation}\label{Eq_H}
    \mathcal{H}(\varphi_a,\varphi_a^*,\varphi_m,\varphi_m^*)=i\frac{\varphi_m}{2}\Big(1-K_L-\frac{3}{4}K_{NL}|\varphi_m|^2-\alpha \gamma_0 \mu -i\xi_m\Big)+i\gamma_0 \mu \frac{\varphi_a}{2}
\end{equation}
The manifold $\mathcal{H}(\varphi_a,\varphi_a^*,\varphi_m,\varphi_m^*)=0$ is called the slow invariant manifold (SIM). This manifold is defined as the fixed points of the dynamical system thanks to equation (\ref{Hyp_asympt}).  As a result, it is the topological space representing the asymptotic solutions of the system.\\
The SIM can be expressed as:
\begin{equation}\label{SIM2}
    \varphi_a=-\frac{\varphi_m}{\gamma_0 \mu}\Big(1-K_L-\alpha \gamma_0 \mu - \frac{3}{4}K_{NL}|\varphi_m|^2 - i\xi_2\Big)
\end{equation}
The complex variables of Manevitch can be expressed in the polar domain:
\begin{equation}
\left\{
    \begin{aligned}
    &\varphi_a=N_a\e^{i\delta_a}\\
    &\varphi_m=N_a\e^{i\delta_m}
    \end{aligned}
\right.
\end{equation}
Replacing the polar domain form of the complex variables of Manevitch in equation (\ref{SIM2}), and calculating the modulus of the SIM using the real and imaginary parts of the equation reads:
\begin{equation}
    N_a=\frac{N_m}{\gamma_0 \mu}\sqrt{\Big(1-K_L-\alpha\gamma_0\mu-\frac{3}{4}K_{NL}N_m^2\Big)^2 + \xi_2^2}
\end{equation}
This equation gives the magnitude of the SIM. In this particular case, the SIM is not a function of phases $\delta_a$ and $\delta_m$.
\subsection{Unstable zones of the SIM}
\label{Unstablezones}
Let us consider the second equation of system (\ref{eps0}) and its complex conjugate:
\begin{equation}\label{SIM3}
    \frac{\partial \varphi_m}{\partial T_0} + i\frac{\varphi_m}{2}\Big(1-K_L-\frac{3}{4}K_{NL}|\varphi_m|^2-\alpha \gamma_0 \mu -i\xi_m\Big)+i\gamma_0 \mu \frac{\varphi_a}{2}=0
\end{equation}
Knowing that $\varphi_a$ is independent of $T_0$, one can search the boundary of unstable zones only perturbing $\varphi_m$:
\begin{equation}
\left\{
    \begin{aligned}
        &\varphi_m \longrightarrow \varphi_m+\Delta\varphi_m\\
        &\varphi_m^* \longrightarrow \varphi_m^*+\Delta\varphi_m^*
    \end{aligned}
\right.
\end{equation}
It gives the following system which can be presented under a matrix form in equation (\ref{matrix_eq}), as we decide to neglect the terms of order $O(\Delta\varphi_m^2)$ and superior orders:
\begin{equation}
\begin{pmatrix}
\frac{\partial\Delta\varphi_m}{\partial T_0}\\
\frac{\partial\Delta\varphi_m^*}{\partial T_0}
\end{pmatrix}
=-\frac{i}{2}
    \mathbf{M}
\begin{pmatrix}
\Delta\varphi_m\\
\Delta\varphi_m^*
\end{pmatrix}
\end{equation}
where the matrix $\mathbf{M}\in \mathbb{M}_{2,2}$ is set as:
\begin{equation}
	\mathbf{M}
	=
	\begin{pmatrix}
		1-K_L-\alpha\gamma_0\mu-\frac{3}{2}K_{NL}|\varphi_m|^2-i\xi_2& -\frac{3}{4}K_{NL}\varphi_m^2\\
		\frac{3}{4}K_{NL}\varphi_m^{*^2} & -\Big(1-K_L-\alpha\gamma_0\mu-\frac{3}{2}K_{NL}|\varphi_m|^2-i\xi_2\Big)
	\end{pmatrix}
\end{equation}
Let us denote the terms of the matrix $\mathbf{M}_{kl}$ with the indices $(k,l)\in\{1,2\}$, where $k$ is the indice of lines and $l$ of columns. The characteristic polynomial $\chi$ of the matrix can be calculated in order to get the eigenvalues:
 \begin{equation}
     \chi(\lambda)=(\mathbf{M}_{11}-\lambda)(\mathbf{M}_{22}-\lambda) - \mathbf{M}_{12}\mathbf{M}_{21}
 \end{equation}
 which yields:
 \begin{equation}
     \chi(\lambda)=\lambda^2 + b\lambda + c
 \end{equation}
 with $b=-\tr (\mathbf{M})=-(\mathbf{M}_{11}+\mathbf{M}_{22})$ and $c=\det (\mathbf{M})=\mathbf{M}_{11}\mathbf{M}_{22}- \mathbf{M}_{12}\mathbf{M}_{21}$.
 
We notice that $(b,c)\in\mathbb{R}^2$, and that $b=\xi_2$. Being a quadratic equation, up to two eigenvalues of the matrix can be obtained, named $\lambda_1$ and $\lambda_2$:
\begin{equation}
    \left \{
    \begin{aligned}
    &\lambda_1=\frac{-b-\sqrt{b^2-4c}}{2}\\
    &\lambda_2=\frac{-b+\sqrt{b^2-4c}}{2}
    \end{aligned}
    \right.
\end{equation}
The unstable and stable manifolds are determined by the sign of the real part of the eigenvalues. Multiple cases arise, depending on the value of the discriminant of the quadratic equation.\\

\textbf{If $b^2-4c<0$}\\
In this case, $(\lambda_1,\lambda_2)\in\mathbb{C}^2$:
\begin{equation}
    \left \{
    \begin{aligned}
    &\lambda_1=\frac{-b-i\sqrt{4c-b^2}}{2}\\
    &\lambda_2=\frac{-b+i\sqrt{4c-b^2}}{2}
    \end{aligned}
    \right.
\end{equation}
It gives that $\Re (\lambda_1)=\Re (\lambda_2)=-\frac{b}{2}=-\frac{\xi_2}{2}$. As the aim of the study is not to inject energy into the system, the damping $\xi_2$ of the ER is taken positive. As a result, the real part of the eigenvalue is negative. This case defines a manifold where the fixed points are stable.\\\\

\textbf{If $b^2-4c=0$}\\
In this case, $(\lambda_1,\lambda_2)\in\mathbb{R}^2$ as:
\begin{equation}
    \lambda_1=\lambda_2=\frac{-b}{2}=-\frac{\xi_2}{2}
\end{equation}
This case defines a manifold where the fixed points are stable.\\

\textbf{If $b^2-4c>0$}\\
In this case, $(\lambda_1,\lambda_2)\in\mathbb{R}^2$:
\begin{equation}
    \left \{
    \begin{aligned}
    &\lambda_1=\frac{-b-\sqrt{b^2-4c}}{2}\\
    &\lambda_2=\frac{-b+\sqrt{b^2-4c}}{2}
    \end{aligned}
    \right.
\end{equation}
It gives the condition $\lambda_1>0 \equiv c<0$. If $c<0$, it defines the unstable manifold, and $c>0$ defines the stable manifold. The condition $c=0$ gives the centre manifold and as a result the boundary between the stable and unstable manifolds.\\
So we can solve the equation $c<0$ to determine the unstable manifold:
\begin{equation}\label{Unstable_cond}
    \frac{1}{4}\Bigg(\xi_2^2 + \Big(1-K_L-\alpha \gamma_0 \mu - \frac{3}{2}K_{NL}|\varphi_m|^2\Big)^2\Bigg)-\Big(\frac{3}{8}K_{NL}\Big)^2|\varphi_m|^4<0
\end{equation}
which can be organized as:
\begin{equation}
    \Big(\frac{27}{16}{K_{NL}}^2\Big)X^2 -3K_{NL}\Big(1-K_{L}-\alpha\gamma_0\mu\Big)X +\Big(\xi_2^2 + (1-K_L-\alpha\gamma_0\mu)^2\Big)<0
\end{equation}
where $X=|\varphi_m|^2=N_m^2$. It is a quadratic equation, which solutions $X_1$ and $X_2$ are:
\begin{equation}\label{sol_unstable}
    \left \{
    \begin{aligned}
        &X_1=\frac{(1-K_L-\alpha\gamma_0\mu)+\frac{1}{2}\sqrt{(1-K_L-\alpha\gamma_0\mu)^2 - 3\xi_2^2}}{\frac{9}{8}K_{NL}}\\
        &X_2=\frac{(1-K_L-\alpha\gamma_0\mu)-\frac{1}{2}\sqrt{(1-K_L-\alpha\gamma_0\mu)^2 - 3\xi_2^2}}{\frac{9}{8}K_{NL}}
    \end{aligned}
    \right.
\end{equation}
Each solution $X_k$, $k=\{1,2\}$ gives two solution for the equation $X_k=N_m^2$. However, only the positive solutions are chosen, as $N_m\in\mathbb{R_+}$. We can observe that in order to get an unstable manifold, the condition $b^2-4c>0$ is needed, which means that $X_1$ and $X_2$ would be two dissociated real solutions. It gives the following condition on the parameters $K_L$ and $xi_2$:
\begin{equation}\label{xi_2c}
    (1-K_L-\alpha\gamma_0\mu)^2 - 3\xi_2^2>0
\end{equation}
This condition in the literature is well known and usually defines a critical value of the damping $\xi_{2,c}$ value. In our study, a critical value of the linear stiffness $K_{L,c}$ can be defined as well. However, the possible range of the value of the linear stiffness is limited by the hypothesis of the loudspeaker being solicited on its linear first mode.
\subsection{Slow system dynamics}
The slow time scale corresponds to the equations at $O(\varepsilon)$. The first equation of (\ref{Eq_Galerkin}) at order $O(\varepsilon)$ gives:
\begin{equation}
    \frac{\partial \varphi_a}{\partial T_1}+\frac{i\varphi_a}{2}(2\sigma -\gamma_0-i\xi_a) + \frac{i\alpha\gamma_0}{2}\varphi_m=\frac{if}{2}
\end{equation}
We can set a manifold defined by $\epsilon(\varphi_a,\varphi_a^*,\varphi_m,\varphi_m^*)=0$, with:
\begin{equation}
    \epsilon(\varphi_a,\varphi_a^*,\varphi_m,\varphi_m^*)=\frac{i\varphi_a}{2}(2\sigma -\gamma_0-i\xi_a) + \frac{i\alpha\gamma_0}{2}\varphi_m-\frac{if}{2}
\end{equation}
Additionally, the evolution of the SIM $\mathcal{H}$ at time scale $T_1$ can be calculated using the chain rule:
\begin{equation}
    \left \{
    \begin{aligned}
        &\frac{\text{d} \mathcal{H}}{\text{d} T_1}=\frac{\partial \mathcal{H}}{\partial \varphi_a}\frac{\partial \varphi_a}{\partial T_1} + \frac{\partial \mathcal{H}}{\partial \varphi_m}\frac{\partial \varphi_m}{\partial T_1} + \frac{\partial \mathcal{H}}{\partial \varphi_a^*}\frac{\partial \varphi_a^*}{\partial T_1} + \frac{\partial \mathcal{H}}{\partial \varphi_m^*}\frac{\partial \varphi_m^*}{\partial T_1}\\
        &\frac{\text{d} \mathcal{H}^*}{\text{d} T_1}=\frac{\partial \mathcal{H}^*}{\partial \varphi_a}\frac{\partial \varphi_a}{\partial T_1} + \frac{\partial \mathcal{H}^*}{\partial \varphi_m}\frac{\partial \varphi_m}{\partial T_1} + \frac{\partial \mathcal{H}^*}{\partial \varphi_a^*}\frac{\partial \varphi_a^*}{\partial T_1} + \frac{\partial \mathcal{H}^*}{\partial \varphi_m^*}\frac{\partial \varphi_m^*}{\partial T_1}
    \end{aligned}
    \right.
\end{equation}
which can be rearranged as:
\begin{equation}
\begin{pmatrix}
\frac{\partial \mathcal{H}}{\partial \varphi_m}&\frac{\partial \mathcal{H}}{\partial \varphi_m^*}\\
\frac{\partial \mathcal{H^*}}{\partial \varphi_m}&\frac{\partial \mathcal{H^*}}{\partial \varphi_m^*}
\end{pmatrix}
\begin{pmatrix}
\frac{\partial \varphi_m}{\partial T_1}\\
\frac{\partial \varphi_m^*}{\partial T_1}
\end{pmatrix}
=-
\begin{pmatrix}
\frac{\partial \mathcal{H}}{\partial \varphi_a}&\frac{\partial \mathcal{H}}{\partial \varphi_a^*}\\
\frac{\partial \mathcal{H^*}}{\partial \varphi_a}&\frac{\partial \mathcal{H^*}}{\partial \varphi_a^*}
\end{pmatrix}
\begin{pmatrix}
\frac{\partial \varphi_a}{\partial T_1}\\
\frac{\partial \varphi_a^*}{\partial T_1}
\end{pmatrix}
\end{equation}
Let $\mathbf{A}$ stand for the jacobian matrix of $(\mathcal{H},\mathcal{H}^*)$ regarding the variables $\varphi_m$ and $\varphi_m^*$:
\begin{equation}
    \mathbf{A}=\begin{pmatrix}
\frac{\partial \mathcal{H}}{\partial \varphi_m}&\frac{\partial \mathcal{H}}{\partial \varphi_m^*}\\
\frac{\partial \mathcal{H^*}}{\partial \varphi_m}&\frac{\partial \mathcal{H^*}}{\partial \varphi_m^*}
\end{pmatrix}
\end{equation}
The method defines the equilibrium and singular points as being the intersection of the manifolds defined by $\epsilon=0$ and $\mathcal{H}=0$, combined with a condition about the invertibility of $\mathbf{A}$. The non-invertibility of $\mathbf{A}$ gives the extremums of the SIM regarding $\varphi_m$ and $\varphi_m^*$ in the time scale $T_1$. As a result, the singular points are defined as:
\begin{equation}\label{Singular}
    \left \{
    \begin{aligned}
        &\epsilon(\varphi_a,\varphi_m,\varphi_a^*,\varphi_m^*)=0\\
        &\mathcal{H}(\varphi_a,\varphi_m,\varphi_a^*,\varphi_m^*)=0\\
        &\det (\mathbf{A})=0
    \end{aligned}
    \right.
\end{equation}
And equilibrium points are defined as:
\begin{equation}\label{Equilibrium}
    \left \{
    \begin{aligned}
        &\epsilon(\varphi_a,\varphi_m,\varphi_a^*,\varphi_m^*)=0\\
        &\mathcal{H}(\varphi_a,\varphi_m,\varphi_a^*,\varphi_m^*)=0\\
        &\det (\mathbf{A})\ne0
    \end{aligned}
    \right.
\end{equation}
\subsubsection{Singular points}
The matrix $\mathbf{A}$ is:
\begin{equation}
    \mathbf{A}=
    \begin{pmatrix}
    \frac{i}{2}(1-K_L-\alpha\gamma_0\mu - \frac{3}{2}K_{NL}|\varphi_m|^2-i\xi_m) &-\frac{3i}{8}K_{NL}\varphi_m^2\\
    \frac{3i}{8}K_{NL}\varphi_m^{*^2} & -\frac{i}{2}(1-K_L-\alpha\gamma_0\mu - \frac{3}{2}K_{NL}|\varphi_m|^2+i\xi_m)
    \end{pmatrix}
\end{equation}
And its determinant is:
\begin{equation}
    \det (\mathbf{A})=\frac{1}{4}\Bigg(\Big(1-K_L-\alpha\gamma_0\mu-\frac{3}{2}K_{NL}|\varphi_m|^2\Big)^2 + \xi_m^2\Bigg) - \frac{9}{64}K_{NL}^2|\varphi_m|^4
\end{equation}
Let us solve the equation $\det (A) =0$:
\begin{equation}\label{detA}
    \frac{1}{4}\Bigg(\Big(1-K_L-\alpha\gamma_0\mu-\frac{3}{2}K_{NL}|\varphi_m|^2\Big)^2 + \xi_m^2\Bigg) - \frac{9}{64}K_{NL}^2|\varphi_m|^4=0
\end{equation}
Equation (\ref{detA}) and equation (\ref{Unstable_cond}) are the same relation, meaning that the singular points coincide with the stability borders of the SIM if the points are on the manifolds defined equation (\ref{Singular}). In this case, the singularities define the stability of the manifold. The solutions of the quadratic equation are given in equations (\ref{sol_unstable}). 
\subsubsection{Equilibrium points}
The equilibrium points are defined by the intersection of the manifolds defined by $\epsilon=0$ and $\mathcal{H}=0$. Let us solve the system defined by equations (\ref{Equilibrium}):
\begin{equation}\label{equilibrium_explicit}
    \left \{
    \begin{aligned}
        &\frac{i\varphi_a}{2}(2\sigma -\gamma_0-i\xi_a) + \frac{i\alpha\gamma_0}{2}\varphi_m-\frac{if}{2}=0\\
        &i\frac{\varphi_m}{2}\Big(1-K_L-\frac{3}{4}K_{NL}|\varphi_m|^2-\alpha \gamma_0 \mu -i\xi_m\Big)+i\gamma_0 \mu \frac{\varphi_a}{2}=0\\
        &\det (\mathbf{A})\ne0
    \end{aligned}
    \right.
\end{equation}
As previously done in equation (\ref{SIM2}), the second equation of system (\ref{equilibrium_explicit}) can be organized to express $\varphi_a$ as a function of $\varphi_m$. This equation can be replaced in the first equation of (\ref{equilibrium_explicit}). The resulting equation is a polynomial equation regarding $\varphi_m$. We express the variable in polar form $\varphi_m=N_m\e^{i\delta_m}$, and we take the modulus of the equation. It becomes a polynomial expression of the third degree:
\begin{equation}\label{eq:pts_equilibre}
    aX^3+bX^2+cX+d=0
\end{equation}
with:
\begin{equation}
    \left \{
    \begin{aligned}
    &X=N_m^2\\
    &a=-\Big(\frac{3}{4}K_{NL}\Big)^2\Big((2\sigma-\gamma_0)^2 + \xi_a^2\Big)\\
    &b=\Big(1-K_L-\gamma_0\mu\alpha\Big)\Big((2\sigma-\gamma_0)^2+\xi_a^2\Big) -\alpha\gamma_0^2\mu (2\sigma-\gamma_0)\\
    &c=\Big(1-K_L-\gamma_0\mu\alpha\Big)^2\Big((2\sigma-\gamma_0)^2+\xi_a^2\Big) + \Big(\xi_a\xi_m+\alpha\gamma_0^2\mu\Big)^2 + \xi_m^2(2\sigma-\gamma_0)^2\\
    &-2\alpha\gamma_0^2\mu\Big(1-K_L-\gamma_0\mu\alpha\Big)\Big(2\sigma-\gamma_0\Big)\\
    &d=-f^2\gamma_0^2\mu
    \end{aligned}
    \right.
\end{equation}
This equation can be solved using the Cardano method. It gives the values of $N_m$ corresponding to the intersection of the two manifolds, which are the equilibrium points or the singularities. The values of $N_m$ placed at the intersection of the manifold are now known, the calculations of the values of $N_a$ can be done using either of the equations defining the manifolds. The condition set by equation (\ref{detA}) gives the nature of the considered points.
\subsection{The case of the experiment}
\subsubsection{Formalization of the equations}
In this study, the nonlinear function chosen to produce a duffing behavior $F_{\text{NL}}(u_m)=K_t\beta_{NL}u_m^3(t)$. Defining the excitation as $p_{ls}(t)=P \sin{(\Omega t)}$ with $\Omega$ its pulsation, system of equation (\ref{EqSystem}) becomes:
\begin{equation}
\left\{
    \begin{aligned}
        &m_a\ddot u_a(t) +c_a \dot u_a(t) + k_a u_a(t)+\frac{\gamma}{\alpha}\big(u_a(t) - \alpha u_m(t)\big)=-S_t P \sin{(\Omega t)}\\
        &M_t \ddot u_m(t)+ R_t \dot u_m(t) + K_t u_m(t) + K_t\beta_{NL}u_m^3(t) +\gamma \big(\alpha u_m(t) - u_a(t)\big)=0 
    \end{aligned}
\right.
\end{equation}
If we set the eigen pulsation of the acoustic first mode of the tube:
\begin{equation}
    \omega_0=\sqrt{\frac{k_a}{m_a}}
\end{equation}
And we introduce the change of variables on the time with the notation $\bullet'=\frac{\partial}{\partial \tau}$:
\begin{equation}
        \tau=\omega_0t
\end{equation}
At the working pressure, we evaluate the order of displacements of the air mass in the reduced section tube and of the membrane of the loudspeaker of the ER. We re-scale the displacements of the system $u_a$ and $u_m$ to a value of order $O(1)$ by introducing the variables $\chi=1\times 10^{-6}$, $u_1$ and $u_2$ with following change of variables:
\begin{equation}
    \left\{
    \begin{aligned}
    &u_a(t)=\chi u_1(t)\\
    &u_m(t)=\chi u_2(t)
    \end{aligned}
\right.
\end{equation}
It yields the system of equations with the normalized pulsation regarding the primary system (the acoustic mode):
\begin{equation}\label{EqSystem2}
\left\{
    \begin{aligned}
        &u_1''(\tau) +\frac{c_a}{\sqrt{k_a m_a}} u_1'(\tau) + u_1(\tau)+\frac{\gamma}{\alpha k_a}\big(u_1(\tau) - \alpha u_2(\tau)\big)=-\frac{S_t P}{k_a\chi} \sin{\Big(\frac{\Omega}{\omega_0} \tau\Big)}\\
        &u_2''(\tau)+ \frac{R_t}{M_t}\sqrt{\frac{m_a}{k_a}} u_2'(\tau) + \frac{K_t m_a}{k_a M_t} u_2(\tau) + \frac{K_t m_a}{k_a M_t}\chi^2\beta_{NL}u_2^3(\tau) +\frac{\gamma}{\alpha k_a}\frac{m_a \alpha}{M_t} \big(\alpha u_2(\tau) - u_1(\tau)\big)=0 
    \end{aligned}
\right.
\end{equation}
In mechanics, the parameter $\varepsilon$ is defined as the ratio between the mass of the nonlinear oscillator and the mass of the primary system. In this case, the mass of the ER is higher than the mass of the air volume. As a result, the parameter $\varepsilon<<1$ is defined here as an arbitrary numerical value.  In order to respect the standard form of equations (\ref{Eq_stand}), one should set the following parameters:
\begin{equation}
    \left\{
    \begin{aligned}
    &\frac{c_a}{\sqrt{k_a m_a}}=\varepsilon \xi_a\\
    &\frac{\gamma}{\alpha k_a}=\varepsilon \gamma_0\\
    &\frac{S_t P}{k_a\chi}=\varepsilon f\\
    &\frac{R_t}{M_t}\sqrt{\frac{m_a}{k_a}}=\xi_m\\
    &\frac{K_t m_a}{k_a M_t}=K_L\\
    &\frac{K_t m_a}{k_a M_t}\chi^2\beta_{NL}=K_{NL}\\
    &\frac{\gamma}{\alpha k_a}\frac{m_a \alpha}{M_t}=\gamma_0\mu\\
    \end{aligned}
\right.
\end{equation}
Where the parameters $\xi_a$, $\gamma_0$, $f$, $\xi_m$, $K_L$, $K_{NL}$ and $\gamma_0\mu$ should be order $O(1)$, and we choose $\varepsilon=10^{-2}$. However, the value of the parameters of the experiment presented in Table \ref{Table_Value} does not allow to respect all the hypotheses. Indeed, the parameter $\gamma_0 \mu$ should be order $O(1)$, which means that the ratio $(m_a \alpha)/M_t$ should be order $O(\varepsilon^{-1})$, due to the previously defined relation $\gamma/(\alpha k_a)=\varepsilon \gamma_0$. The value of $(m_a \alpha)/M_t=0.42$ does not respect the hypothesis made in the analytical developments due to the weak coupling. We set $\mu$ as $\varepsilon (m_a \alpha)/M_t=\mu$.\\
For simplification purposes, the targeted parameters of the ER are chosen by the choice of the coefficients $\eta_M$, $\eta_R$ and $\eta_K$ as:
\begin{equation}
    \left\{
    \begin{aligned}
    &M_t=\eta_M M_0\\
    &R_t=\eta_R R_0\\
    &K_t=\eta_K K_0
    \end{aligned}
\right.
\end{equation}
The rescaled dimensionless parameters are presented in Table (\ref{Table_rescaled}), with the choice of $\beta=5.0\times10^{10}\ \text{m}^{-2}$, $\eta_R=1/8$, and $\eta_M=\eta_K=1$, and the value of the parameters $M_0$, $R_0$ and $K_0$ presented in Table \ref{table_ER_0}. The choice of $\eta_R$ is driven by the condition of equation (\ref{xi_2c}).\\
\begin{table}[ht!]
    \centering
    \begin{tabular}{c|c}
    $\xi_a$     &$13.05$\\
    $\gamma_0$   &$6.71$\\
    $\xi_m$  &$2.49\times10^{-2}$\\
    $K_L$    &$9.65\times10^{-1}$\\
    $K_{NL}$   &$4.82\times10^{-2}\ \text{m}^{-2}$\\
    $\mu$   & $4.16\times10^{-3}$
    \end{tabular}
    \caption{Values of the rescaled parameters}
    \label{Table_rescaled}
\end{table}
As a result, the system of equations is organized as the standard form equations (\ref{Eq_stand}), and the method presented can be applied, while one should keep in mind the limits of the hypothesis made here.
\subsubsection{Analytical Results}
\begin{figure}[htp] 
	\centering
	\subfloat[$\beta_{\text{NL}}=5\times10^{10}\ \text{m}^{-2}$, $\eta_K=0.9$]{%
		\includegraphics{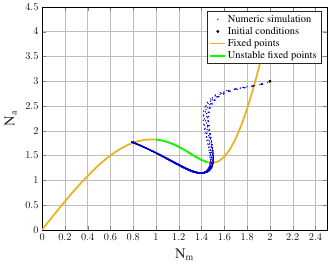}%
		\label{Figure SIM_hard}%
	}\hfil
	\subfloat[$\beta_{\text{NL}}=-5\times10^{10}\ \text{m}^{-2}$, $\eta_K=0.9$]{%
		\includegraphics{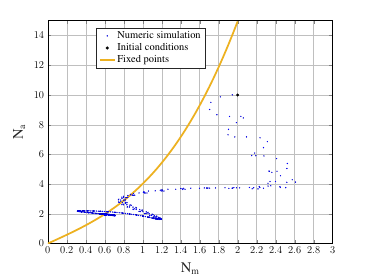}%
		\label{Figure_SIM_soft_uk0_9}%
	}
	
	\subfloat[$\beta_{\text{NL}}=-5\times10^{10}\ \text{m}^{-2}$, $\eta_K=1.222$]{%
		\includegraphics{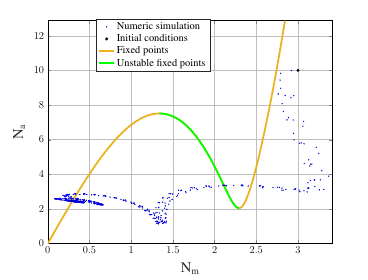}%
		\label{Figure_SIM_soft_uk1_222}%
	}\hfil
	\subfloat[$\beta_{\text{NL}}=5\times10^{10}\ \text{m}^{-2}$, $\eta_K=0.9$]{%
		\includegraphics{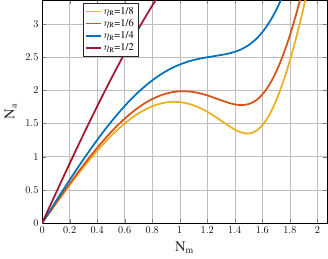}%
		\label{fig:Bifur_etaR}%
	}
	\caption{Figures of the SIM of the two degree of freedom system defined by equation (\ref{SIM2}), with $P=0.5$ Pa}
	\label{fig:SIMs}
\end{figure}
Let us study the results given by the analytical calculations. Each analytical results are supported by numerical simulations. Multiple phenomena can be anticipated using analytic and numeric approaches. The SIM presented equation (\ref{SIM2}) with the stability boundary conditions introduced equation (\ref{sol_unstable}) is depicted in Figure \ref{Figure SIM_hard} with the given parameters and $\eta_K=0.9$. The unstable fixed points whose existence indicates eventual bifurcations are indicated in green. In fact, an additional condition about $\eta_K$ can be observed: if the linear frequency of the ER is not tuned with the frequency of the acoustic mode of the tube, the bifurcation can not happen. It is due to the coupling between the pressure inside the reduced-section tube and the ER, such as the pressure needs to reach the activation threshold of the nonlinear behavior of the ER. Indeed, the linear frequency of the ER should be chosen such as the nonlinear resonance is aligned with the acoustic mode of the tube. Taking $\eta_K=0.9$ lowers the linear frequency of the ER which aligns the nonlinear resonance with the resonance of the reduced-section tube since the behavior is hardening. This effect can also be observed by taking an interest in the softening behavior. If we set $\beta_{\text{NL}}=-5\times10^{10}\ \text{m}^{-2}$ and we keep $\eta_K=0.9$, the bifurcation can not happen as presented in Figure \ref{Figure_SIM_soft_uk0_9}. However, if we set $\eta_K=1.222$, the linear frequency is placed at higher frequencies than the frequency of the reduced-section tube resonance, leading to the alignment of the nonlinear resonance of the ER with the frequency of the reduced-section tube resonance. It causes the existence of unstable fixed points, resulting in possible bifurcations as depicted in Figure \ref{Figure_SIM_soft_uk1_222}. Additionally, the condition of existence of unstable zones given equation (\ref{xi_2c}) is verified in Figure \ref{fig:Bifur_etaR}. In fact, the critical damping parameter with the considered parameters is $\xi_{2,c}=0.0444$, which yields $\eta_R=0.2232$. It can be seen that the SIM with $\eta_R=1/4$ does not include unstable zones, while the SIMs for $\eta_R=1/8$ and $\eta_R=1/6$ do. 
\begin{figure}[htp] 
	\centering
	\subfloat[$\beta_{\text{NL}}=5\times10^{10}\ \text{m}^{-2}$, $\eta_K=0.9$]{%
		\includegraphics{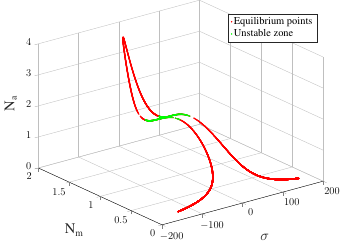}%
		\label{fig:pts_eq_hard}%
	}\hfil
	\subfloat[$\beta_{\text{NL}}=-5\times10^{10}\ \text{m}^{-2}$, $\eta_K=1.222$]{%
		\includegraphics{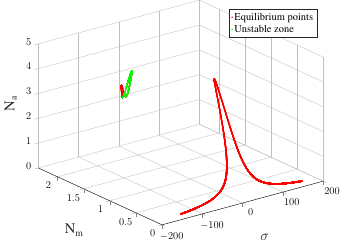}%
		\label{fig:pts_eq_soft}%
	}
	\caption{Figures of the equilibrium points of the two degree of freedom system defined by equation (\ref{eq:pts_equilibre}), with $P=0.8468$ Pa}
	\label{fig:Pts_EQ}
\end{figure}
Let us plot the equilibrium points of the two degrees of freedom system with both the softening and hardening restoring forces. We solve the equation (\ref{eq:pts_equilibre}) by finding the roots of the polynomial expression using the Cardano method. The results are plotted in Figure \ref{fig:Pts_EQ} with incident pressure $P=0.8486$ Pa which is the numerical input that suits the experimental pressure input.
\begin{figure}[htp] 
	\centering
	\subfloat[$\beta_{\text{NL}}=5\times10^{10}\ \text{m}^{-2}$, $\eta_K=0.9$]{%
		\includegraphics{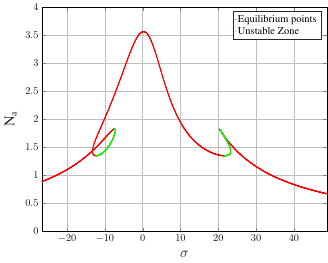}%
		\label{Pt_eq_hard_Na}%
	}\hfil
	\subfloat[$\beta_{\text{NL}}=5\times10^{10}\ \text{m}^{-2}$, $\eta_K=0.9$]{%
		\includegraphics{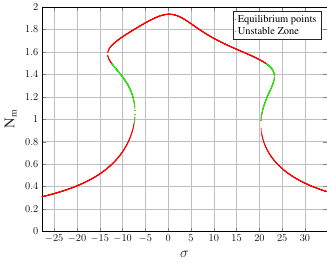}%
		\label{Pt_eq_hard_Nm}%
	}
	
	\subfloat[$\beta_{\text{NL}}=-5\times10^{10}\ \text{m}^{-2}$, $\eta_K=1.222$]{%
		\includegraphics{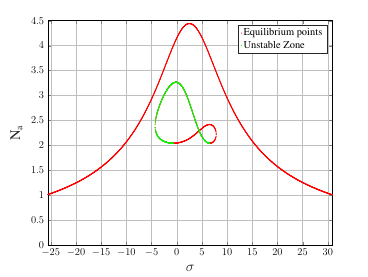}%
		\label{Pt_eq_soft_Na}%
	}\hfil
	\subfloat[$\beta_{\text{NL}}=-5\times10^{10}\ \text{m}^{-2}$, $\eta_K=1.222$]{%
		\includegraphics{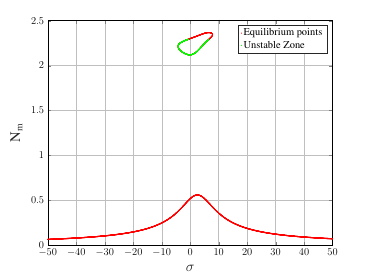}%
		\label{Pt_eq_soft_Nm}%
	}
	\caption{Figures of the equilibrium points of the two degree of freedom system defined by equation (\ref{eq:pts_equilibre}), with $P=0.8468$ Pa}
	\label{fig:EQ_2D}
\end{figure}
One can observe in Figure \ref{fig:Pts_EQ} that both behaviors do not demonstrate quasi-periodic regime with initial conditions set to 0. It is needed to pre-stress to system to attain quasi-periodic regimes. Moreover, possible bifurcations can be expected from the system with hardening behavior with the given parameters due to the unstable zones of the main branch, as it can be seen in Figure \ref{fig:pts_eq_hard}. Additionally, the Figure \ref{Pt_eq_hard_Na} predicts that a bifurcation may happen around $\sigma=20$. Regarding the softening case in Figure \ref{fig:pts_eq_soft}, we observe an isolated branch of equilibrium points that is far from the main branch. However, Figure \ref{Pt_eq_soft_Na} shows that the isolated branch is situated at lower energy than the main branch for the acoustic mode, and at higher energy than the main branch for the ER (Figure \ref{Pt_eq_soft_Nm}). Indeed, the larger the amplitudes of the ER, the lower the amplitudes of the acoustic mode, as the ER can absorb more energy with large oscillations. 

\section{Experimental results}
\label{Sec3}
\subsection{Setup}
\begin{figure}[htp] 
	\centering
	\includegraphics[scale=0.15]{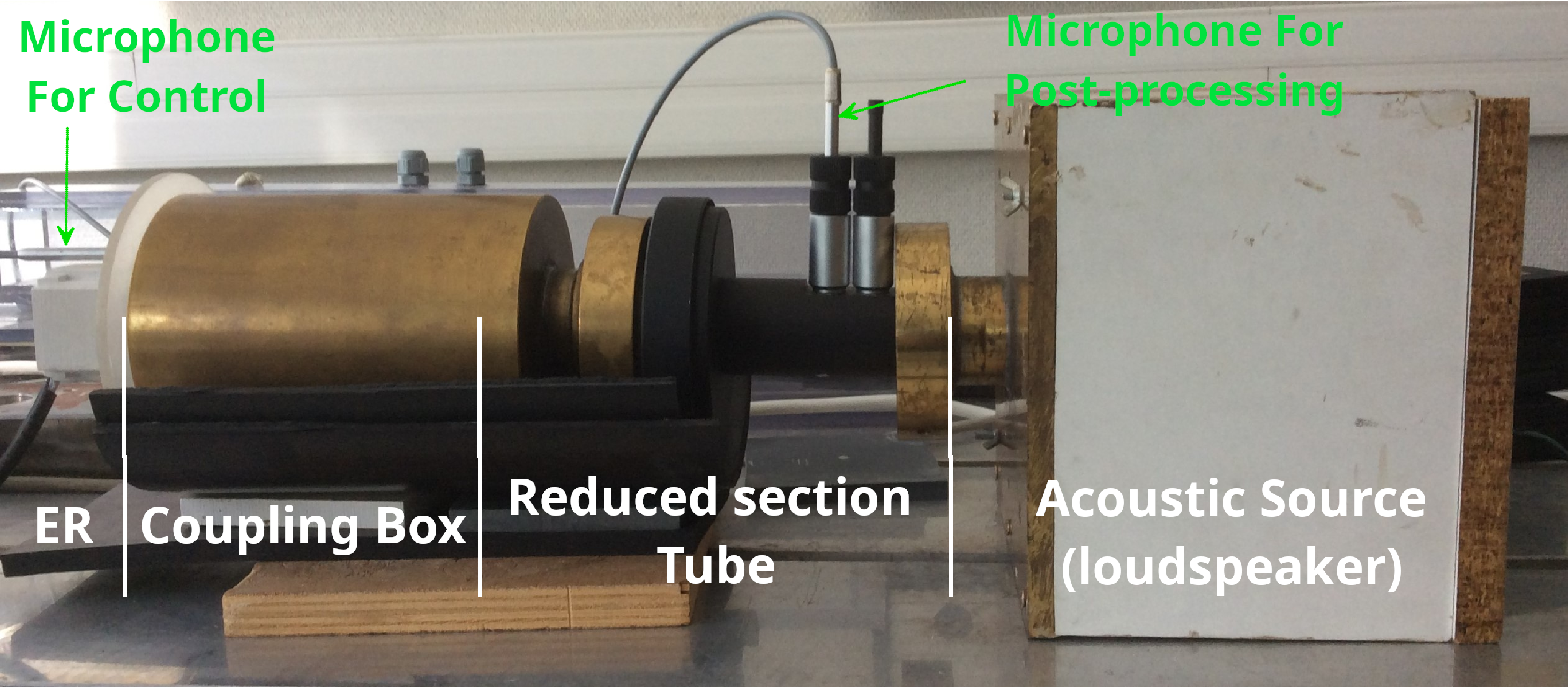}
	\caption{Experimental set-up}
	\label{fig:Exp_setup}
\end{figure}

The experimental setup is pictured in Figure \ref{fig:Exp_setup}. The dimensions and parameters have been presented in Tables \ref{table_ER_0}, \ref{Table_Value}, \ref{Param_calculated} and \ref{Param_measured}. The digital programming of the ER is done using a D-Space MicroLabBox DS1202, which is also used as the acquisition system working at 50 kHz. The electrical current is not amplified between the controlling device and the ER as the calculated electrical current should be equal to the electrical current to be injected into the loudspeaker coil. The ER is power-supplied using a 10 V amplitude tension. The excitation is provided by an external loudspeaker linked to the D-Space device through an amplifier. Reference measures for all considered excitation signals are made using a rigid termination instead of the ER.\\
In this section, both hardening and softening behavior are presented. For the experiments, the specified parameters for the hardening behavior are kept as $\eta_M=\eta_K=1$, $\beta_{\text{NL}}=5\times10^{10}$. The targeted parameters for the softening behaviors are $\eta_M=1$ $\eta_K=1.222$, $\beta_{\text{NL}}=-5\times10^{10}$. Multiple nonlinear phenomena will be shown, such as faster decrease in transient regime, or the condition of existence of the bifurcations. Bifurcations will be showcased, and a discussion about the potential of such system for noise reduction is realized based on the experimental results.

\subsection{Results}
\subsubsection{Validation of the model by numerical simulation} \label{Simulations}
\begin{figure}[ht!]
	\centering
	\subfloat{\includegraphics{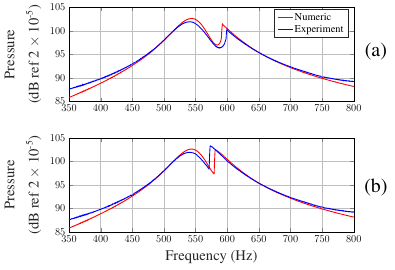}}
	\subfloat{\includegraphics{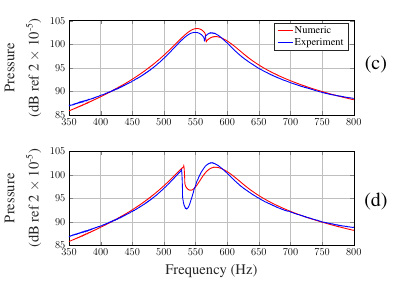}}
	\caption{Experimental and corresponding numerical results obtained from direct numerical integration of Eq. (\ref{EqSystem}): variation of the pressure amplitude for the hardening behavior with (a) increasing frequency sweep, (b) decreasing frequency sweep and for the softening behavior with (c) increasing frequency sweep, (d) decreasing frequency sweep.}
	\label{Numeric}
\end{figure}

This section aims at validating the analytical model. To that end, the system of equations (\ref{EqSystem}) is solved using an order 4 Runge-Kutta method. The numeric excitation signal is taken as identical to the experimental one, i.e. a frequency sweep from 350 Hz to 800 Hz that lasts 60 seconds with either increasing or decreasing frequency sweeps. Variations of pressure inside the reduced-section tube for both experimental and numerical results with increasing and decreasing frequency sweeps are presented in Figure \ref{Numeric}. The parameters are $\eta_K=1$, $\beta_{\text{NL}}=5\times10^{10}\ \text{m}^{-2}$ for the hardening behavior and $\eta_K=1.222$, $\beta_{\text{NL}}=-5\times10^{10}\ \text{m}^{-2}$ for the softening behavior. Notice that these parameters are the ones used for the experiment.
It can be observed that the numerical integration of the analytical model is very efficient in predicting the behavior of both hardening and softening behaviors with either increasing or decreasing frequency sweeps. The frequency of the bifurcation is also well-predicted in all cases. Nevertheless, one can observe that the absorption of the ER for the softening behavior with the decreasing frequency sweep (Figure \ref{Numeric}d) is better for the experiment than for the numerical simulation. It is caused by the intrinsic resonance of the ER loudspeaker which is close to the cavity resonance to control. Indeed, the intrinsic resonance of the ER loudspeaker is not considered in the reduced order model, and it is also not deleted by the programming. This improved absorption of the experimental data with respect to the numerical data has to be remembered to evaluate the efficiency of the analytical study done with the MSM coupled to the complex variables of Manevitch method.

\subsubsection{The hardening and softening cubic behavior frequency response function}
\begin{figure}[htp] 
	\centering
	\subfloat[With hardening restoring force function]{%
		\includegraphics{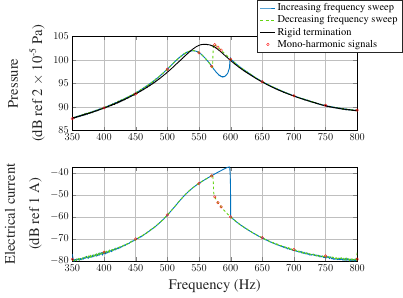}%
		\label{fig:CubHard}%
	}\hfil
	\subfloat[With softening restoring force function]{%
		\includegraphics{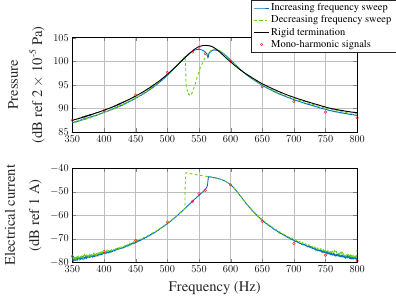}%
		\label{fig:CubSoft}%
	}
	\caption{Variations of pressure inside the reduced-section tube and electrical current injected in the ER under frequency sweeps excitation}
	\label{fig:Cubique_sweep}
\end{figure}

In this section, the nonlinear cubic restoring force implemented in an ER is studied with multiple excitation signals. Increasing and decreasing frequency sweeps are considered, as well as mono-harmonic signals. The frequency sweeps last 60 seconds and start at 350 Hz and end at 800 Hz. Mono-harmonic signals are composed of a 5-second long sinus, while the acquisition system also records the decrease in transient regime. The amplitude of mono-harmonic signals is retrieved and is plotted on the Function Response Function (FRF). Both hardening and softening behaviors are experimentally implemented. Variations of pressure inside the reduced-section tube and electrical current injected in the ER under frequency sweeps excitation are presented in Figure \ref{fig:Cubique_sweep}. One can observe that bifurcations occur as brutal changes of equilibrium points can be seen. Moreover, a frequency bandwidth where two stable points of equilibrium exist is showcased by the increasing and decreasing frequency sweeps. Additionally, the pressure in the reduced-section tube is significantly reduced when the ER behaves with large oscillations, which are placed at its nonlinear resonance. The oscillations are maximized in the case of a hardening behavior for increasing frequency sweep, and in the case of a softening behavior for decreasing frequency sweep. However, it can be seen that mono-harmonic signals excite the lower branch in amplitude of the resonator, resulting in a higher pressure in the reduced-section tube. These observations mean that if the ER is not pre-stressed, i.e. has initial conditions without energy, the ER will select the lower branch of its equilibrium points. Therefore, the decreasing frequency sweep for the hardening behavior and the increasing frequency sweep for the softening behavior are equivalent to the response under mono-harmonic signals if the resonator is not pre-stressed. Nevertheless, the excitation signals that maximize the ER response, i.e. increasing frequency sweep for hardening behavior and decreasing frequency sweep for softening behavior could be obtained with pre-stressed resonators. Future studies will explore the possibility of using the higher energy branch of the resonator to maximize the energy transfer from the primary system (the acoustic mode of the tube) to the nonlinear resonator. \\
\begin{figure}[htp] 
	\centering
	\subfloat[$\beta_{\text{NL}}=5\times10^{10}\ \text{m}^{-2}$, $\eta_K=1.122$]{%
		\includegraphics{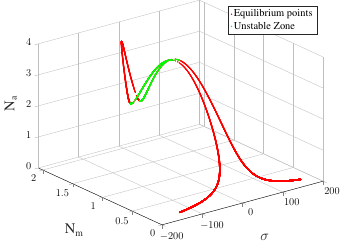}%
		\label{Pt_eq_soft_muK1_122}%
	}
	\subfloat[$\beta_{\text{NL}}=5\times10^{10}\ \text{m}^{-2}$, $\eta_K=1.122$]{%
		\includegraphics{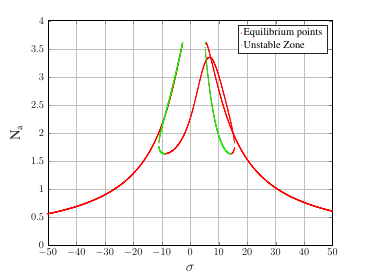}%
		\label{Pt_eq_soft_muK1_122_Na}%
	}\hfil
	\subfloat[$\beta_{\text{NL}}=-5\times10^{10}\ \text{m}^{-2}$, $\eta_K=1.122$]{%
		\includegraphics{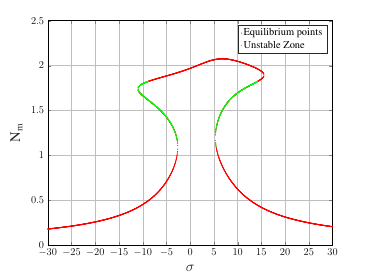}%
		\label{Pt_eq_soft_muK1_122_Nm}%
	}
	\caption{Figures of the equilibrium points of the two degree of freedom system defined by equation (\ref{eq:pts_equilibre}), with $P=0.8468$ Pa}
	\label{fig:EQ_corr}
\end{figure}

One can observe that the analytical results predict well the shape of the curves in the case of the hardening behavior, such as the Figure \ref{Pt_eq_hard_Na}. The bifurcation is well localized (after the resonance), and towards higher amplitude. Notice that the equilibrium points obtained by analytical means describe a mono-harmonic response, and do not describe the response to the frequency sweep signal. However, the analytical results indicated that $\eta_K=1$ does not allow unstable zones to exist. This shift can be explained by the hypothesis made during the analytical development on the scale of the coupling parameter. Additionally, the differences that we can observe between analytical and experimental data, such as the value of the parameter $\eta_K$, are probably not caused by the intrinsic resonance of the ER loudspeaker which is close to the cavity resonance to control as previously mentioned. Indeed, this discrepancy does not seem to be caused by the model as it can be seen from the numerical simulations of Section \ref{Simulations}. The parameters used for the numerical simulations are the parameters of the experiment. The same phenomena can be seen for the softening behavior, as the parameter $\eta_K=1.222$ does not describe well the experimental results. Choosing the parameter $\eta_K=1.122$ to plot analytical results (i.e. taking the same shift of -0.1 as for the hardening behavior) gives the Figure \ref{fig:EQ_corr}. A more accurate description of the phenomena is observed.

\subsubsection{The transient regime}
\begin{figure}[htp] 
	\centering
	\subfloat[Transient decrease from the higher branch of the ER (frequency sweep signal)]{%
		\includegraphics{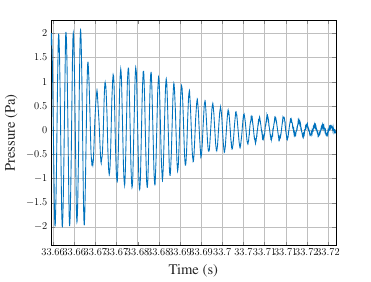}%
		\label{fig:High580}%
	}\hfil
	\subfloat[Transient decrease from the lower branch of the ER (mono-harmonic signal)]{%
		\includegraphics{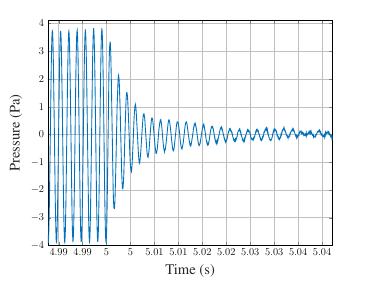}%
		\label{fig:Sin580}%
	}
	\caption{Pressure inside the reduced-section tube at 580 Hz}
	\label{fig:Cubique_transitoire}
\end{figure}
To study the transient regime, two different sets of measures were carried out with two different excitation signals. The transient decrease analysis is done from a mono-harmonic signal featuring a 5-second long sinus, and from a frequency sweep signal. The frequency sweep signal starts at 350 Hz and stops on a chosen frequency for 3 seconds. This excitation signal has been designed to enable the study of the transient regime while the ER state is placed on the higher amplitude branch. The two decreases from Figure \ref{fig:Cubique_transitoire} can be compared in terms of shape. The transient decrease of the primary system when the ER is placed on a high energy equilibrium point is presented in Figure \ref{fig:High580}. It highlights a first decrease, and then an increase in amplitude can be noted. It leads to an exponential decrease until an amplitude of 0 Pa. The transient decrease of the primary system when the ER is placed on a low energy equilibrium point is presented in Figure \ref{fig:Sin580}. Only an exponential decrease can be observed. The different phenomena are due to either the energy level of the ER is higher or lower than the bifurcation minimum energy.

\subsubsection{Condition of existence of the bifurcations}
\begin{figure}[htp] 
	\centering
	\subfloat[$\eta_R=\frac{1}{8}$]{%
		\includegraphics{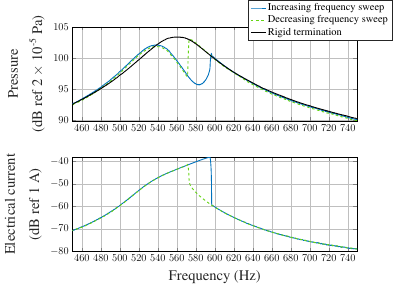}%
		\label{fig:1_8}%
	}\hfil
	\subfloat[$\eta_R=\frac{1}{6}$]{%
		\includegraphics{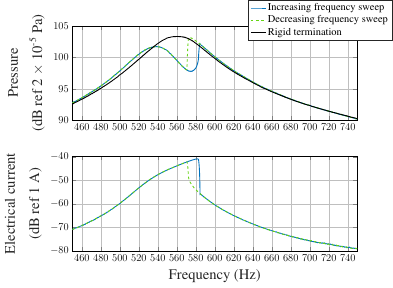}%
		\label{fig:1_6}%
	}
	
	\subfloat[$\eta_R=\frac{1}{4}$]{%
		\includegraphics{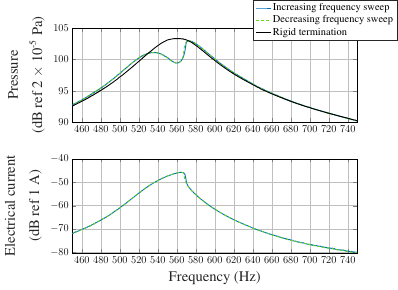}%
		\label{fig:1_4}%
	}\hfil
	\subfloat[$\eta_R=\frac{1}{4}$]{%
		\includegraphics{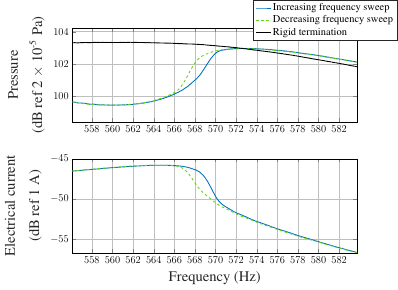}%
		\label{fig:1_4_zoomed}%
	}
	
	\subfloat[$\eta_R=\frac{1}{2}$]{%
		\includegraphics{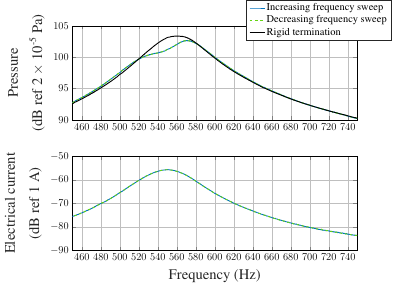}%
		\label{fig:1_2}%
	}\hfil
	\subfloat[$\eta_R=\frac{1}{2}$]{%
		\includegraphics{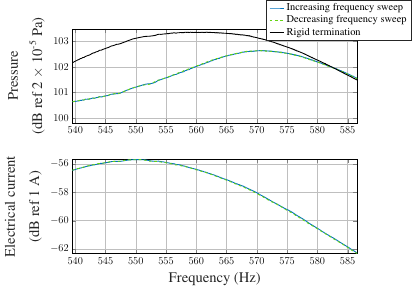}%
		\label{fig:1_2_zoomed}%
	}
	
	\caption{Variations of pressure inside the reduced-section tube and electrical current injected in the ER under frequency sweeps excitation for different values of damping $\eta_R$}
	\label{fig:Condition_Existence}
\end{figure}
As presented in Section \ref{Unstablezones}, the bifurcation only exists within a range of the resonator damping values. To study this condition, the system is excited using a frequency sweep with both increasing and decreasing frequency sweeps. Let us consider the hardening cubic behavior with a fixed linear stiffness such as $\eta_K=1$, and multiple damping values $\eta_R$. The critical damping value calculated using the experimental data is $\eta_{R,c}=0.3244$. Higher values of $\eta_R$ prevent the bifurcations to occur. Variations of pressure inside the reduced section tube and electrical current injected into the loudspeaker's coil for various values of the damping parameter $\eta_R$ are presented in Figure \ref{fig:Condition_Existence}. One can observe that the analytical results about the role of damping on the existence of bifurcations are accurate. Figures \ref{fig:1_8} and \ref{fig:1_6} show that bifurcations occur at lower values of damping $\eta_R=1/8$ and $\eta_R=1/6$ than the critical value $\eta_{R,c}$. Moreover, the frequency bandwidth of the existence of at least two stable equilibrium points is narrowed with higher $\eta_R$. From the analytical study, it is known that the unstable zone of the SIM becomes smaller with the rise of the damping value $\eta_R$. Nevertheless, bifurcations are smoothed when $\eta_R$ approach the critical value $\eta_{R,c}$ as seen in Figures \ref{fig:1_4} and \ref{fig:1_4_zoomed}, which is due to the two stable equilibrium points getting very close in amplitudes. When the damping is superior to the critical damping, the bifurcations do not exist, and there is not a frequency bandwidth where two stable equilibrium points exist, as seen in Figures \ref{fig:1_2} and \ref{fig:1_2_zoomed}. The agreement with analytical results is quite good, as the critical damping value is well predicted.

\section{Conclusion}
The paper focuses on the predictions of experimental results using an analytical method to study nonlinear systems. The Multiple Scales Method coupled to the complex variables of Manevitch lies on multiple assumptions, which are not fully respected here. It is shown that the method is suitable to study the phenomena and have a better understanding of them. The critical damping value can be well predicted, and the behavior of the two degrees of freedom system can be qualitatively predicted. However, the inaccuracy of the prediction of the quantitative values is due to the model that is not complete. Indeed, it fails to describe the intrinsic resonance of the ER loudspeaker which is not deleted by the programming. As a result, despite the numerous assumptions needed by the analytical method and the reduced order model, its predictions can be used to understand the phenomena at stake.

\section*{Acknowledgement}
The authors would like to thank the following organizations for supporting this research: (i) The "Minist\`ere de la transition \'ecologique" and (ii) LABEX CELYA (ANR-10-LABX-0060) of the "Universit\'e de Lyon" within the program "Investissement d\text{'}Avenir" (ANR-11- IDEX-0007) operated by the French National Research Agency (ANR).


\bibliographystyle{unsrt}
\bibliography{library_publi_7}

\end{document}